\documentclass[multphys,vecphys]{svmult}

\usepackage{makeidx}     % allows index generation
\usepackage{graphicx}    % standard LaTeX graphics tool
\usepackage{subfigure}   % when including figure files
\usepackage{multicol}    % used for the two-column index
\makeindex             % used for the subject index
\begin{document}

\title*{Methanol: a diagnostic tool for high-mass star-forming regions}
% Use \titlerunning{Short title} for an abbreviated version of
\titlerunning{Methanol excitation}% for an abbreviated version of
\authorrunning{Leurini et al.}
\author{S. Leurini \inst{1}, P. Schilke\inst{1}, K.M. Menten\inst{1},
  D.R. Flower\inst{2}, J.T. Pottage\inst{2}
  \and L.-H Xu\inst{3}}
\institute{Max-Planck-Institut f\"ur Radioastronomie, Auf Dem H\"ugel 69, D-53121, Bonn
\texttt{sleurini@mpifr-bonn.mpg.de}
\and Physics Department, University of Durham, DH1 3LE, UK 
\and Physical Sciences Department, University of New Brunswick, Saint John, NB, Canada E2L 4L5}
\maketitle
\section{Introduction}\label{intro}
Kinetic temperature and density are fundamental parameters for our
understanding of the interstellar medium (ISM).
 Usually, symmetric rotors such as NH$_3$ are used
to probe a cloud's kinetic temperature, while linear molecules, e.g.
CS, probe its density.  However, different spatial distributions of
the tracers (``chemistry'') often complicate the picture
(see, e.g., \cite{tafalla}) 
as they often trace physically different
and spatially non-coexisting gas components. It is thus desirable to
trace all relevant physical parameters with a single molecule.
Promising candidates exist among slightly asymmetric rotors, which
have properties qualifying them as tracers for physical conditions.
Methanol, CH$_3$OH, is a slightly asymmetric rotor. It is
ubiquitous and associated with different regimes of star formation,
from quiescent, cold ($\mathrm T \sim 10$ K), dark clouds,  
to ``hot core'' sources near high-mass (proto)stellar objects, where
[CH$_3$OH/H$_2$]
values $\sim 10^{-7}-10^{-6}$ are observed \cite{menten}.
Up to now an extremely poor knowledge of the CH$_3$OH
collisional rates and of their propensity rules has prevented
realistic systematic studies exploiting methanol's full potential as
an interstellar tracer. Recently, this situation has changed with the
calculation of collisional rate coefficients by
\cite{pottage1,pottage2}, for collisions with helium, for both
CH$_3$OH-$A$ and CH$_3$OH-$E$, for levels up to
$(\mathrm J,\mathrm K)=9$.\\
Here we would like to focus on general aspects connected to the analysis 
of complex molecules' spectra, of which CH$_3$OH is one of the simplest examples, 
(for details on 
methanol excitation and on its probing properties see \cite{leurini}). 
\section{Analysis technique}\label{an}
\begin{figure}[htbp]
\centering                                         
\includegraphics[width=8cm]{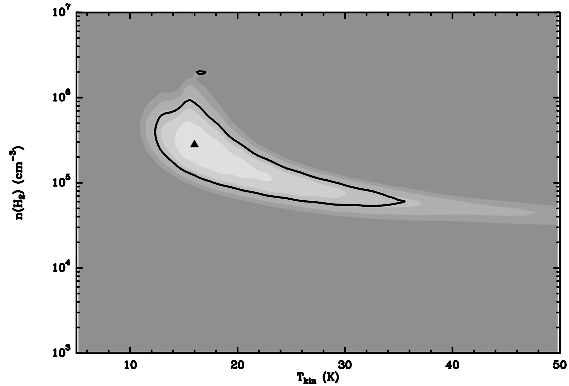}
\caption{ $\chi^2$ distribution in the [T$_{kin}$,n(H$_2$)] plane, N(CH$_3$OH-A)=7.9 10$^{13}$,
  N(CH$_3$OH-E)=5.7 10$^{13}$, toward G79.3P1. 
A black triangle marks the minimum position, T$_{kin}$=17~K, n(H$_2$)=2.5 10$^5$ cm$^{-3}$; 
dashed in black the 3$\sigma$ confidence surface for 22 degrees of freedom.}\label{chi}
\end{figure}

\begin{figure}[htbp]
\centering
\includegraphics[width=12cm]{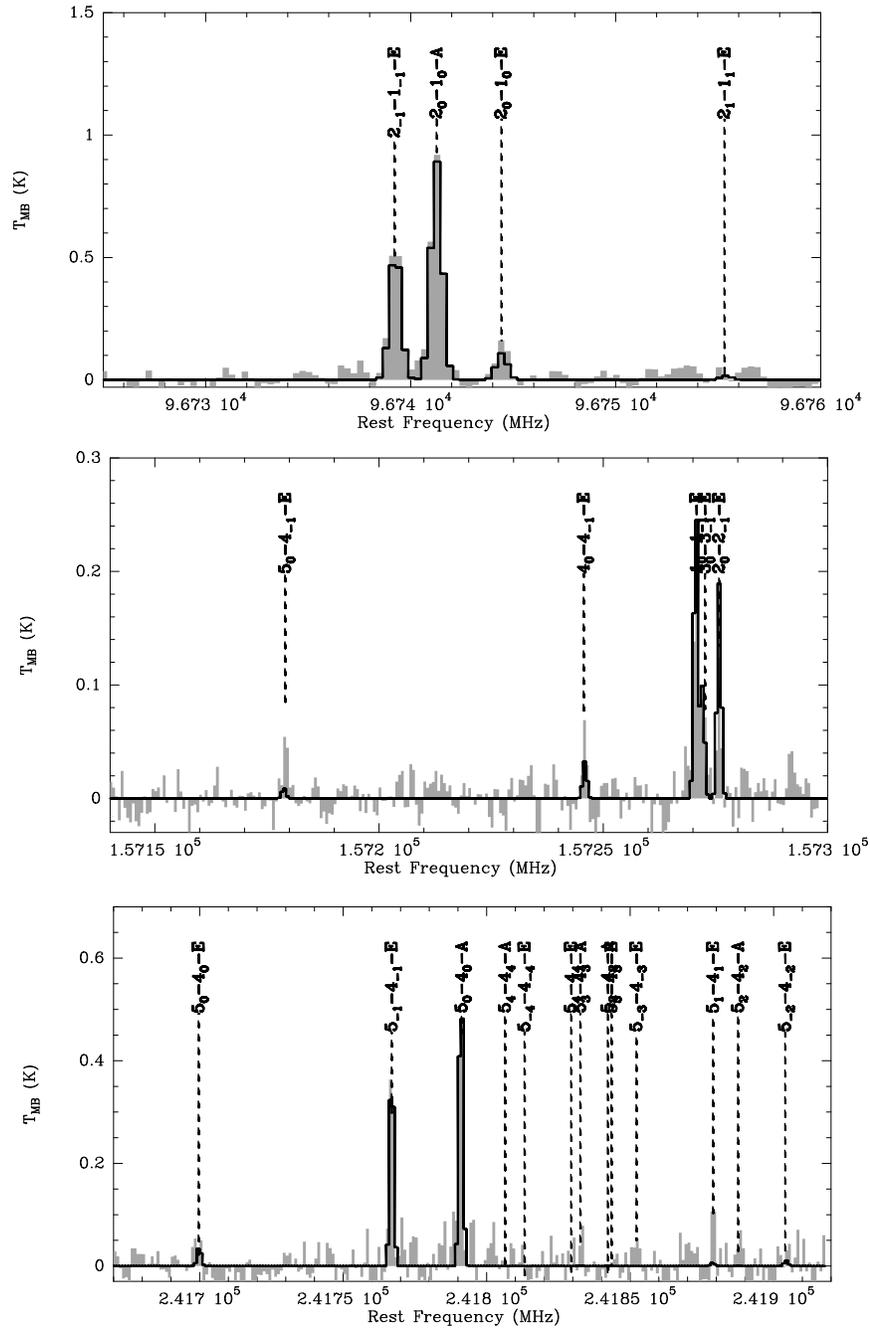}
\caption{CH$_3$OH spectra toward G79.3P1 taken with the IRAM 30m telescope; the 2mm and 1mm data have been 
smoothed to the 3mm beam size ot the IRAM telescope for comparison with the 3mm data. 
Labels show all the CH$_3$OH transitions. Overlaid in black the synthetic spectrum 
resulting from the fit.}\label{fitdata3}
\end{figure}

The traditional approach for deriving physical parameters such as
kinetic temperature and spatial density from an observed spectrum
involves ``by-hand'' fitting of the lines with multiple components and
$\chi$$^2$ analysis comparing the measured quantities with statistical
equilibrium calculations.
However, this single line fitting procedure is extremely 
time-consuming and suffers from several drawbacks.
With sensitive receivers nowadays
available and in view of the next generation of instruments, which will
provide copious amounts of data in a short time, new methods of data analysig 
and modelling are required.\\
An innovative technique to handle the problem, proposed by
\cite{schilke} and recently improved by \cite{comito},
is based on the simultaneous fit of the complete spectrum with a
synthetic spectrum computed under Local Thermodynamic Equilibrium
(LTE) conditions. Here we propose an extension to this technique using
the Large Velocity Gragient (LVG) approximation, which ensures more reliable results when 
prominent departures from LTE are expected.
The free parameters
for each component are excitation temperature, molecular hydrogen
density, source size, column densities for A and E states, which are
treated as  two independent parameters to take into account 
their possibly different abundance.
Line width and LSR velocity are fixed parameters
and assumed to be the same for all the lines in each component.
Within a component,
the optical depths for lines with a frequency separation 
$\nu_{i}- \nu_{j} \le \Delta\nu_{i}+\Delta\nu_{j}$ 
are summed up to include local line overlap between both symmetry states. 
Line identification is based on the Cologne Database for Molecular 
Spectroscopy, 
(http://www.cdms.de \cite{muller}), which includes new measurements 
by \cite{xu}.\\
Although the simultaneous fit of a spectrum has been demonstrated to
be a powerful technique of analysis
 \cite{comito}, it does imply several approximations.
The assumption behind the analysis is that the CH$_3$OH distribution can be reasonably
well approximated by a small number of non-interacting
components, thus ignoring any structure in the source. 
The LVG approach adds other uncertainties to the obtained results, since it assumes
only one set of physical parameters for the source and implicitly
neglects any local and non-local overlap between the lines. 
 Also, fitting CH$_3$OH spectra assures a
reliable determination of kinetic temperature only when a large
amount of lines are fitted and/or when millimeter and submillimeter
data are combined, since CH$_3$OH shows a strong dependence 
on temperature mainly in the 
submillimeter range \cite{leurini}. 
Another drawback is the uniqueness
of the $\chi$$^2$ minimization, which is not \emph{a priori}
guaranteed. Indeed the fit results occasionally depend on the
input parameters: an analysis of the $\chi$$^2$ distribution is therefore 
necessary to verify that the found minimum is global.\\   
Having all these limitations in mind, the application of this technique to 
G79.3P1, an infrared-dark cloud, 
gives very satisfying results, (Fig.~\ref{fitdata3}).
Table~\ref{modeltable} lists the best fit results 
and the $3\sigma$ fit range. The source size, $32''$, 
is based on a BIMA map of 
CH$_3$OH at 3mm, (Wyrowski, priv. comm.).
Fig.~\ref{chi} shows the $\chi^2$ distribution in the [T$_{kin}$,n(H$_2$)] plane, 
with the 3$\sigma$ confidence surface in black. 
\begin{table*}[hbt]
\centering
\caption{CH$_3$OH model results for G79.3P1, $\chi_{\nu_{d}}^2$=0.7, $\nu_{d}$=22.\label{modeltable}}
\begin{tabular}{lcc}
\hline
\hline
&best fit&3$\sigma$ fit range\\
T$_\mathrm K$&17~K&12--37~K\\
n(H$_2$)&\multicolumn{1}{r}{2.5 10$^5$(cm$^{-3})$}&
\multicolumn{1}{r}{3~$ 10^5$--~4~$10^6$(cm$^{-3})$}\\
N(CH$_3$OH-A)&\multicolumn{1}{r}{7.9 10$^{13}$ (cm$^{-2})$}&
\multicolumn{1}{r}{6~$10^{13}$--1~$10^{14}$(cm$^{-2})$} \\
N(CH$_3$OH-E)&\multicolumn{1}{r}{5.7 10$^{13}$ (cm$^{-2})$}&
\multicolumn{1}{r}{3~$10^{13}$--8~$10^{13}$(cm$^{-2})$}\\
\hline
\hline
\end{tabular}
\end{table*}
%\bibliographystyle{aa} 
%\bibliography{bibsilvia}
% BibTeX users please use
% \bibliographystyle{}
% \bibliography{}
%
% Non-BibTeX users please follow the syntax
% the syntax of "referenc.tex" for your own citations
%%%%%%%%%%%%%%%%%%%%%%%% referenc.tex %%%%%%%%%%%%%%%%%%%%%%%%%%%%%%
% sample references
% "physics"
%
% Use this file as a template for your own input.
%
%%%%%%%%%%%%%%%%%%%%%%%% Springer-Verlag %%%%%%%%%%%%%%%%%%%%%%%%%%

%\input{referenc.tex}
%%%%%%%%%%%%%%%%%%%%%%%%%%%%%%%%%%%%%%%%%%%%%%%%%%%%%%%%%%%%%%%%%%%%%%  }

%%%%%%%%%%%%%%%%%%%%%%%%%%%%%%%%%%%%%%%%%%%%%%%%%%%%%%%%%%%%%%%%%%%%%%

\printindex
\end{document}